**Multi-wavelength and multi-messenger Fast Radio Burst follow-up**

Sarah Burke-Spolaor

Center for Gravitational Waves and Cosmology, West Virginia University, Chestnut Ridge Research Building, Morgantown, WV 26505. Email: sarah.spolaor@mail.wvu.edu

Department of Physics and Astronomy, West Virginia University, Morgantown, WV 26506, USA

*Multi-wavelength and multi-messenger astronomy will reveal the phenomena that produce Fast Radio Bursts, turning Fast Radio Bursts into sharper tools with which to probe extragalactic plasma.*

So far, Fast Radio Bursts (FRBs) are exclusively a radio-wave phenomenon, but a considerable amount of information is contained in the radio pulse. Radio-wavelength light is greatly affected by intervening plasma (see Fig. 1), and so direct measurables from a radio burst itself provide a rich data-set with which to probe both the emission mechanism of these intense flashes (for instance, one can limit the scale of the emitter based on the duration of the burst and light travel time arguments), and all intervening media between the burst source and Earth (e.g. using dispersion, scattering, and Faraday rotation effects to infer something about the intervening medium – see Table 1). This includes information about ionization in the source, any related host galaxy, the intergalactic medium, the circumgalactic medium, and our own Milky Way. One major aim of FRB science is to use these pieces of information to probe baryonic matter, turbulence scales, and polarization in the intergalactic medium. However, there are limitations to the use of radio data: the intergalactic medium can only be studied well if we know the *distance* to the FRB well, and this requires first the identification of a host galaxy through precise FRB localization via radio interferometry, followed by an optical spectra or photometry of that host. We then will need characterization of the different media along the line-of-sight by looking at the ionization, polarization, and turbulence properties of the host galaxy and any FRB-local material. For this, observations at multiple wavelengths and timescales are needed, and this is where multi-wavelength and multi-messenger astronomy are essential.

Thus, while FRBs are unique in their capability as physical probes, we will necessarily need to rely on FRB localization, followed by studies of the source and host with other wavelengths and messengers, to disentangle FRB-local and intergalactic effects. If FRBs are detectable as multi-messenger or multi-wavelength phenomena, those follow-up observations will also be the most direct way to answer the simple (but yet unanswered!) question: "what makes FRBs?"

Gravitational waves, neutrinos, and the full electromagnetic spectrum are now open for business, and time-domain astronomy is reaping the benefits of this multi-wavelength and multi-messenger era. Observing the sequential appearance of light in various wavebands has allowed us to unpack the cause-and-effect evolution of dynamic phenomena such as pulsars, supernovae, gamma-ray bursts, and active galactic nuclei. Our ability to detect both neutrinos and gravitational radiation has added an extra dimension to this field; while the bulk of electromagnetic emission traces the movements of electrons and plasma, neutrinos can inform us about energetic atomic decay processes and hadronic accelerations. Meanwhile, gravitational waves directly track the movement of mass in explosive and/or relativistic phenomena.

So far, there have been only two *multi-messenger* transient sources of cosmic origin: first, a binary neutron star merger in galaxy NGC 4993 was detected by its gravitational-wave and broadband electromagnetic emissions by LIGO/Virgo. This transient event has exemplified the physical detail that we can get from multi-messenger analysis, and has given us an incontrovertible link between binary neutron stars, gamma-ray bursts, kilonovae, and the evolutionary tract that governs energy outflows in this merger phenomenon[1]. Second, the recent detection of a high-energy neutrino from active galaxy TXS 0506+056 has observationally linked blazars as an accelerator of extragalactic-origin cosmic rays.

A multi-messenger detection of an FRB would serve several purposes. The repeating FRB121102 has demonstrated that not all FRBs are cataclysmic, and it is expected that the most conclusive indicator of a cataclysmic subclass of FRB sources would be through the detection of gravitational waves. The associated wave form would reveal critical information about the nature of that progenitor. Similarly, the detection of neutrinos associated with an FRB (and in particular the timing of FRB vs. neutrino detection) would be indicative of the nature and evolution of a cataclysmic progenitor for FRBs.

In Table 1, we provide a summary of the information that could be gleaned from various observations, including the information we can gather from the radio FRBs themselves. When we begin to detect FRBs with multiple messengers and multiple frequencies, those observations will add key time-resolved evolution and dynamical information about the local environment; in an ideal situation, we could use this information to separate the relative influences of the progenitor, host, and IGM. For instance, if we can measure the luminosity and expansion of shock fronts in media around the burst source, we can estimate the progenitor's local electron density, and thereby remove it from the total dispersion contribution.

So, what's stopping us?

Well, knowing what messengers to expect from FRBs relies on the characteristics of the (yet unknown) progenitors, the (not yet directly identified) environments, and the (only roughly constrained) distances to FRBs. The one FRB with a known host galaxy, the repeating FRB121102, has shown only an association with a lone persistent radio source with minimal variability[2]. Currently, the most popular interpretation of that object is a young, highly magnetized neutron star. For pulsars, we consider the "spin-down luminosity" (the rate of energy release based on the observed slowing of a pulsar's rotation) as the available energy budget for any emissions seen. In young magnetars, one can maintain a spin-down luminosity that exceeds that of normal pulsars by more than four orders of magnitude. Therefore, through some mechanism—perhaps via large magnetic flares—this energy reservoir can be drawn upon to make the intense bursts we view as FRBs[3].

With a neutron star source, one would not necessarily expect many neutrinos in the absence of significant hadronic acceleration. Observable gravitational waves *may* be detectable, particularly if FRB-producing magnetars less than a few tens of Mpc away have a sufficiently large spin-down energy and surface deformities ("mountains") greater than a few meters high (the sensitivity of ground-based gravitational wave experiments limits their detection volume to include only a few hundred standard pulsars in our Galaxy). Other proposed cataclysmic progenitors of (non-repeating) FRBs provide more natural routes to multi-messenger emissions of the neutrino or gravitational-wave variety: e.g. neutron star mergers, supramassive neutron star collapse, or cosmic string cusps.

It is not yet clear whether there are multiple types of FRB progenitor (see the Comment by Ue-Li Pen in this issue), and if any among those progenitors will turn out to be anything but radio-emitters. Because of this and the large locational uncertainties for FRBs detected to date, thus far follow-up searches for FRBs other than FRB121102 have represented a full-ocean fishing expedition.

We do have some basic limitations as to what signatures might accompany FRBs. For instance, we know that dense outflows are unlikely to exist concurrently with FRBs due to the fact that we see FRBs at all; if FRBs are within too dense a plasma, the GHz-frequency emissions would be absorbed, and thus not seen. However, this is one of very few true limitations we have on the source(s) of FRBs. To demonstrate why FRB follow-up efforts are currently so widely spread in both wavelength and timescales, consider Table 2, in which we outline just a few of the potential signatures that might accompany (only a small portion of the) many objects that have been proposed to produce FRBs or FRB-like events.

The first FRB detections that were identified in real-time—and subsequently followed up—were localized to a sky region of about the angular scale of the full moon. Even with a rough redshift cut based on an FRB's dispersion (Table 1), for most FRBs this still leaves tens to hundreds or more of candidate host galaxies that could be potentially associated with the event. This fact led these early follow-up programs to survey broadly for "anything that moved"—that is, any object in the error region with detectable variability at any wavelength or timescale[4–6]. Statistically, one would expect variance at some wavelength given enough time and a large enough area; thus naturally, interpreting this data to find conclusive counterpart associations proved difficult[5,7]. This same issue applies for attempts to associate FRBs with a past event in archival data (as has been attempted for neutrinos, gravitational waves, gamma-ray bursts, and supernovae).

However, these studies were able to rule out specific progenitor source classes: for instance, rapid-response observations with Swift indicated that FRB140514 did not have a direct relationship with long-gamma-ray-burst events, while slower radio and optical follow-up did not detect the afterglows that would have accompanied a superluminous supernova remnant out to the nominal $z \lesssim 3$ redshift limit of this FRB[4].

On the multi-messenger front, there have been explicit searches for both gravitational waves and for neutrinos that are temporally and spatially coincident with FRB events. Virgo and GEO 600 data have been searched for gravitational waves coincident with 14 FRBs[8], while twelve and four

FRBs occurred during the operation of the ANTARES and IceCube neutrino telescopes, respectively[9,10]. There have not yet been any positive detections of FRB associations with either of these messengers.

It is worth pointing out that radio telescopes can see only a small portion of the sky, while neutrino and gravitational-wave detectors are by nature omni-directional. As we've estimated that approximately one FRB occurs somewhere in the sky every eight seconds, constraints on the net event rates from these detectors on specific source classes may well be more constraining than individual associations; in fact, this argument was recently used to demonstrate that FRBs could not all be due to relativistic mergers, because the rate of FRBs in the sky is nearly two orders of magnitude too high[11].

Looking to the near future, this field is set to develop rapidly, driven by new facilities. Experiments are now operational that can perform precise FRB localizations in real-time, allowing us to place more focused resources on single-galaxy targets. The Realfast detector on the Very Large Array was able to identify the host of FRB121102 with sub-arcsecond localization[2], while CHIME, ASKAP, and others will soon begin to achieve sub-arcminute localization. All of these facilities are working to put into place automatic trigger alert systems, and there is an ongoing effort to define a standard VOEvent format for FRB event releases[12].

In the next 3–5 years, we expect at least a few if not tens of FRBs to be associated with a host galaxy, and the nature of FRB progenitors and science will unfold during this time. In the meantime, vast multi-wavelength and multi-messenger coordination has now been set in place by hand for various projects, as exemplified by the "Deeper Wider Faster" survey. This program performs targeted FRB search campaigns, but includes pre-coordinated observations at multi-wavelength observatories. With coordinated observation, this program can seek transients that might precede or follow the emission of an FRB. Its scope is far-reaching (Figure 2), and demonstrates the full gamut of telescopes that we are now using to chase FRBs not only across the electromagnetic spectrum, but across all available messengers.

*Sarah Burke Spolaor is at West Virginia University, Morgantown, WV, USA.*
*Email: sarahbspolaor@gmail.com*

References
1. Mooley, K. P. et al. *Nature* https://doi.org/10.1038/s41586-018-0486-3 (CE: not in an issue yet)
2. Chatterjee, S. et al. *Nature* **541**, 58-61 (2017).
3. Metzger, B. D., Berger, E.& Margalit, B. *Astrophys. J.* **841**, 14 (2017).
4. Petroff, E. et al. *Mon. Not. R. Astron. Soc.* **447**, 246 (2015).
5. Murase, K., Mészáros, P & Fox, D. B. *Astrophys. J.* **836**, 6 (2017).
6. Bhandari, S. et al. *Mon. Not. R. Astron. Soc.* **475**, 1427 (2018).
7. Williams, P. K. G. & Berger, E. *Astrophys. J.* **821**, 22 (2016).
8. Abbott, B. P., *Phys. Rev. D* **93**, id.122008 (2016).
9. Albert A. et al., arXiv:1807.04045 (2018).

10. Fahey, S. et al. *Astrophys. J.* **845**, 14 (2017).
11. Callister, T., Kanner, J. & Weinstein, A. *Astrophys. J.* **825**, 12 (2016).
12. Petroff, E. et al. arXiv:1710.08155 (2017).

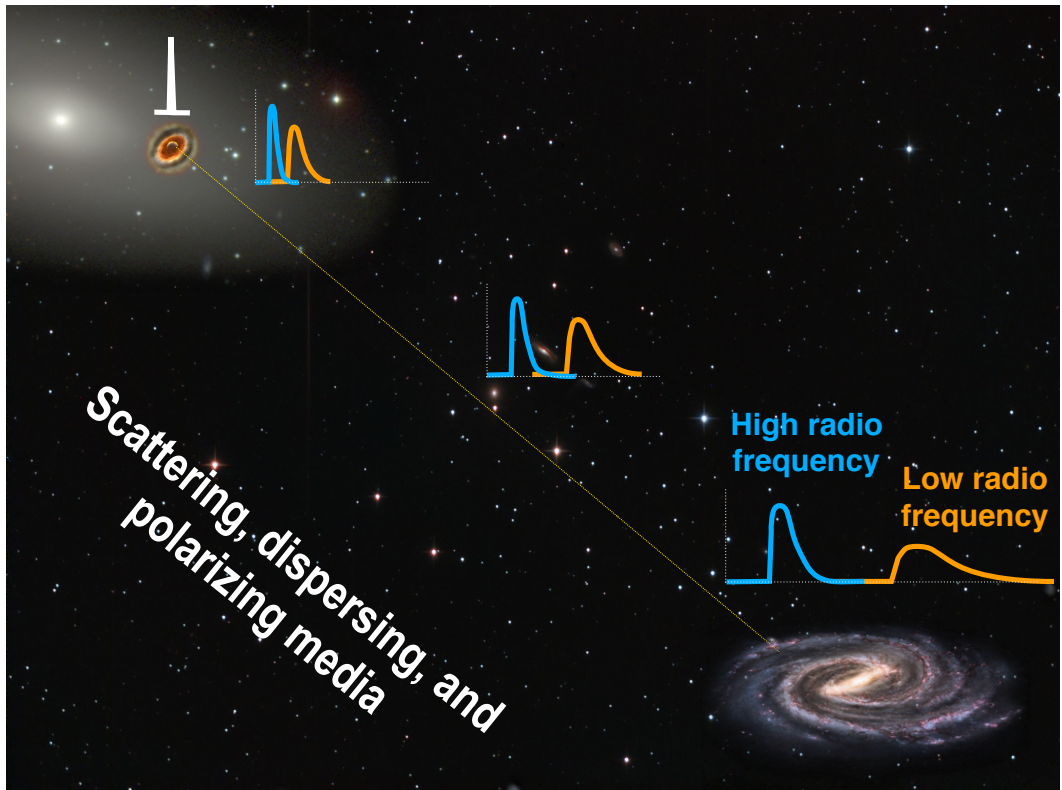

*Figure 1:* **FRBs can uniquely probe the intergalactic medium via propagation effects, however the signals include the effects of the progenitor and any host galaxy.** If we detect multiple messengers or wavelengths from an FRB source, we can extract essential facts about the host environment. In this Figure we are showing the effect of dispersion (a time-dependent sweep from low to high frequency) and scattering (a greater broadening of the pulse at low frequencies). Both of these effects become more pronounced with signal propagation over longer spatial scales.

**Table 1. Information derived from observations of FRB sources.**

| Observation | Measurement | What is inferred |
|---|---|---|
| **FRB** | Dispersion Measure | Integral electron density, distance |
| | Rotation Measure | Integral electron density, distance, B-field |
| | Pulse scattering timescale | Scatterer turbulence and distance |
| | Scintillation bandwidth | Scatterer turbulence and distance |
| | Scattering index | Intervening medium turbulence |
| | Polarization fraction/angle | Magnetic field evolution, Faraday effects |
| | Spectrum | Emission mechanism, optical depth |
| **Electromagnetic Follow-up** | Redshift | Distance |
| | Existence of co-located emission | Presence of local material, star formation |
| | Time evolution | Presence and evolution of afterglow, shocks, jets |
| | Broad-band continuum | Emission process, age, density of any nebula |
| **Gravitational-wave Detection** | Existence of signal | Cataclysmic or cyclic underlying process |
| | Time evolution | Direct dynamics of material |
| **Neutrino Detection** | Existence of signal | Hadronic accelerations; possible presence of beaming |

**Table 2.Multi-wavelength and multi-messenger counterparts of possible FRB sources**. Red text represents radio emission, green represents optical through UV emission, blue represents X-rays, purple represents gamma-rays, and brown represents gravitational wave counterparts.

| Progenitor | Precursor: years to decades | Precursor: < years | Concurrent emission | Seconds-hours | Days | Months | Years |
|---|---|---|---|---|---|---|---|
| Young pulsar/magnetar: 10-100 years old | Long GRB; Superluminous Supernova | | | Faint(?), decaying residual pulsar wind nebula →; Repeat FRBs with high but declining rotation measure → | Repeating radio bursts →; Declining dispersion measure → | | |
| Young pulsar/magnetar: < years old | — | Gravitational-wave Burst | | Evolving multi-wavelength afterglow → | | | |
| Black hole outburst | — | — | Repeat FRBs; Periodic variation of rotation measure and polarization angle →; Persistent radio source →; Extreme UV accretion disk →; X-ray accretion disk → | | | | |
| Blitzar | — | — | Short GRB; No FRB repeat | X-ray afterglow → | | | |
| Black Hole Evaporation | — | — | Multi-wavelength transient? | — | | | |
| Stellar-mass black hole coalescence | — | — | Gravitational-wave Burst; Short GRB | — | | | |
| Superradiance from primordial black holes | | | | Single or multiple FRBs →; Single or multiple gravitational waves → | | | |

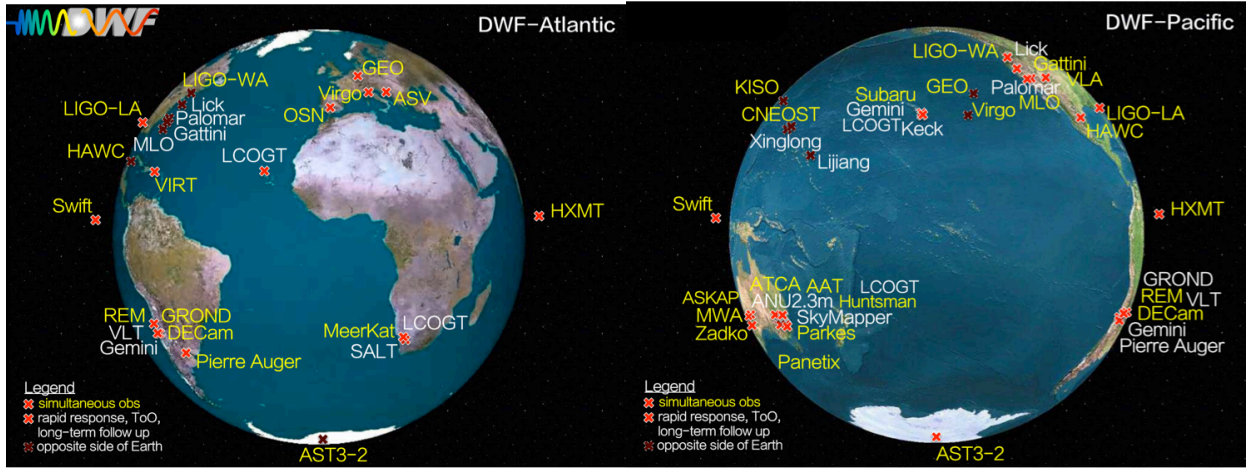

*Figure 2:* **The telescopes involved in the worldwide "Deeper Wider Faster" program.** This effort aims to detect multi-wavelength and multi-messenger counterparts to FRBs and other transients. It seeks FRB precursor, concurrent, and any related afterglow emission through coordinated simultaneous observations, and rapid-response follow-up imaging and spectroscopy. Radio telescope facilities in the network with real-time FRB detection capabilities drive the observations and triggers. Image credit: J. Cooke/Swinburne University of Technology.